\documentclass[journal]{IEEEtran}

\usepackage{graphicx}
\usepackage{hyperref}
\usepackage{amsmath}
\usepackage{amssymb}
\usepackage{mathtools}
\usepackage{adjustbox}
\usepackage{lipsum}
\usepackage{algorithm}
\usepackage{graphicx}
\usepackage{textcomp}
\usepackage{mathtools}
\usepackage{adjustbox}
\usepackage{lipsum}
\usepackage{stackengine}
\usepackage{subfig}
\usepackage{graphicx}

\usepackage{algpseudocode}
\usepackage{expl3}
\usepackage{float}
\usepackage[utf8x]{inputenc}

\begin{document}

\title{ProductGraphSleepNet: Sleep Staging using Product Spatio-Temporal Graph Learning with Attentive Temporal Aggregation}

\author{Aref Einizade, Samaneh Nasiri, Sepideh Hajipour Sardouie, and Gari Clifford
\thanks{Aref Einizade, and Sepideh Hajipour Sardouie are with the Department of Electrical Engineering, Sharif University of Technology, Tehran, Iran. Samaneh Nasiri is with the Massachusetts General Hospital, Harvard Medical School, Georgia, USA. Gari Clifford is with the Georgia Institute of Technology, Georgia, USA, and Emory School of Medicine, Georgia, USA.}}

\markboth{}
{Shell \MakeLowercase{\textit{et al.}}: Bare Demo of IEEEtran.cls for IEEE Journals}
\maketitle

\begin{abstract}
The classification of sleep stages plays a crucial role in understanding and diagnosing sleep pathophysiology. Sleep stage scoring relies heavily on visual inspection by an expert that is time-consuming and subjective procedure. Recently, deep learning neural network approaches have been leveraged to develop a generalized automated sleep staging and account for shifts in distributions that may be caused by inherent inter/intra-subject variability, heterogeneity across datasets, and different recording environments. However, these networks ignore the connections among brain regions, and disregard the sequential connections between temporally adjacent sleep epochs. To address these issues, this work proposes an adaptive product graph learning-based graph convolutional network, named ProductGraphSleepNet, for learning \textit{joint} spatio-temporal graphs along with a bidirectional gated recurrent unit and a modified graph attention network to capture the attentive dynamics of sleep stage transitions. Evaluation on two public databases: the Montreal Archive of Sleep Studies (MASS) SS3; and the SleepEDF, which contain full night polysomnography recordings of 62 and 20 healthy subjects, respectively, demonstrates performance comparable to the state-of-the-art (Accuracy: 0.867;0.838, F1-score: 0.818;0.774 and Kappa: 0.802;0.775, on each database respectively). More importantly, the proposed network makes it possible for clinicians to comprehend and interpret the learned connectivity graphs for sleep stages. 
\end{abstract}


\IEEEpeerreviewmaketitle

\section{Introduction}
Abnormal sleep is increasingly recognized as a crucial factor in many illnesses \cite{wulff2010sleep}. Specifically, sleep physiology recorded via polysomnography (PSG) provides a rich source of information about the brain and cardiovascular health \cite{lajnef2015learning}. Therefore, automated sleep staging and diagnostics of sleep disorders are likely to play a crucial role in large-scale epidemiological research linking sleep patterns to disease and wellness applications \cite{wulff2010sleep}. The ground truth for sleep staging remains the multi-lead electroencephalogram (EEG), where sleep experts use the standard rules (e.g., Rechtschaffen and Kales (R\&K) \cite{wolpert1969manual} and the American Academy of Sleep Medicine (AASM) \cite{iber2007aasm}) for annotating sleep stages. The rules focus on 30-sec windows of data (or `epochs’) and manual labeling of epochs into five stages: Wake (W), Rapid Eye Movement (REM), Non-REM 1-3 (N1, N2, and N3).  In addition to the time and cost involved in manual sleep staging, the significant inter-expert subjectivity may lead to a noisy-labeling issue \cite{jia2020graphsleepnet}. 

To address the aforementioned issues, neural networks have received much attention for developing an automated sleep staging network using physiological time series. In particular, convolutional neural networks (CNNs) have been exploited to handle the multimodality of PSG signals \cite{chambon2018deep,perslev2021u} and to extract subject-invariant representation across the population. Due to the time-series nature of PSG signals, hybrid networks, the combination of convolutional and recurrent layers, have been shown to be the most successful networks for sleep staging task \cite{seo2020intra, sun2019hierarchical, dong2017mixed}. These networks extract informative features to capture the dynamics of sleep stage transitions \cite{supratak2017deepsleepnet, phan2019seqsleepnet}. More specifically, the key advantage of hybrid networks is that they do the sleep staging task in an end-to-end learning framework, removing (almost) all pre-processing steps, which include cleaning data, extracting hand-crafted features, and selecting important features. For example, U-Sleep \cite{perslev2021u}, DeepSleepNet \cite{supratak2017deepsleepnet} and SeqSleepNet \cite{phan2019seqsleepnet} were proposed to handle varying combinations of the PSG signals, different lengths sleep epochs, and capture the temporal dependencies between neighboring epochs. However, since the inputs of these networks are on regular grids (e.g., image-like), these networks ignore the connections among brain regions. Therefore, such networks cannot capture the brain's geometrical information, which can be best described in a non-Euclidean space \cite{jia2020graphsleepnet, barachant2011multiclass}, and they disregard the connected sequential nature of PSGs encoding information about the dynamics of sleep stage transitions. These limitations make it difficult for clinicians to interpret the network's outputs.

To handle and better interpret graph structured data, graph neural network (GNN) and graph convolutional network (GCN) models have been proposed, which provide performing node-level, edge-level, and graph-level prediction tasks \cite{wu2020comprehensive, velivckovic2017graph, zhang2018link}, and have been used for sleep staging tasks \cite{jia2020graphsleepnet, ahmedt2021graph}. Graph convolution operations can be performed in a spatial domain (the space of the nodes) \cite{velivckovic2017graph,niepert2016learning,hamilton2017inductive} or a spectral domain (eigenvalue decomposition of a Graph Shift Operator, e.g. adjacency or Laplacian matrix) \cite{bruna2013spectral,defferrard2016convolutional,levie2018cayleynets}. Wu et al. \cite{wu2020comprehensive} has provided a survey on GCNs using spatial and spectral approaches. Note that although GCN-based models can achieve excellent performance in different domains, they usually use prior known or fixed graphs to perform graph convolution operations \cite{wu2020comprehensive}. This assumption is not optimal in many time-series classification tasks, such as sleep stage classification, since the connectivity graph can change adaptively with the stage transitions \cite{jia2020graphsleepnet}. To the best of our knowledge, GraphSleepNet \cite{jia2020graphsleepnet} is the only application of adaptive graph learning for performing automated sleep staging. However, the authors ignored the weights between the sequential epochs' interactions. In addition, the \textit{joint} attentive spatio-temporal information connections was not captured, and the extracted information from neighbor (sequential) temporal epochs were simply concatenated. To address these issues, we propose an adaptive GCN, named ProductGraphSleepNet, which exploits Graph Signal Processing (GSP) \cite{ortega2018graph} and Product Graph Learning (PGL) \cite{kadambari2020learning} concepts to \textit{jointly} learn the sparsest possible spatio-temporal graph representations.

Training and testing of our method is performed in a cross-subject manner on two well-known and public sleep datasets Montreal Archive of Sleep Studies (MASS)-SS3 \cite{o2014montreal} and SleepEDF database \cite{goldberger2000physiobank, kemp2000analysis}, which contain full night PSG recordings of 62 and 20 healthy subjects, respectively. Specifically, we present three contributions in this work as:

\begin{itemize}
	\item Jointly interpretting and learning of temporal and spatial graphs for each sleep stage to provide a more medically interpretable sleep staging network and model any possible connection between the temporal information. Note that in the GraphSleepNet work \cite{jia2020graphsleepnet}, the spatial graphs are learned using the information of the target epochs only, and temporal information is simply concatenated with the same weight, therefore, ignoring the importance of connection weights between sequential epochs. 
	\item Utilizing bi-directional gated recurrent units (BiGRU) to learn the transition rules between the sleep stages and produce feature vectors for the temporal graphs' nodes.
	\item Adaptively learning the importance of temporal weights for sequential epochs by modifying the graph attention network (GAT) \cite{velivckovic2017graph} module, resulting to proposing a Graph-wise Attention Network (GwAT).
\end{itemize}

\section{Preliminaries}
\label{Prel}

\textbf{Notation}. For indexing the data elements or row/column arrays, we use the MATLAB indexing system, that is, \(\bold{A}(i,j)\), \(\bold{A}(i,:)\) and \(\bold{A}(:,i)\) denote the \((i,j)\)th element, the \(i\)th row and the \(i\)th column of \(\bold{A}\), respectively. In a 3D array \(\bold{A}\), \(\bold{A}(i,:,:)\) means the \(i\)th slice of the first dimension of \(\bold{A}\). The all-zero, all-one vectors, and also the trace operator of \(\bold{A}\) are stated as \(\bold{0}\), \(\bold{1}\), and \(tr(\bold{A})\), respectively.

Let \(\bold{S}_i \in \mathbb{R}^{Q\times T_s}\) represent the \(i\)th 30-sec sleep epoch of the neighbor (sequential) epochs \(\bold{S}\), and the PSG signals of the sequential sleep epochs are stated as \( \bold{S}=(\bold{S}_{t-d},…,\bold{S}_{t},…,\bold{S}_{t+d}) \in \mathbb{R} ^ {P \times Q \times T_s }\), where \(t\) indicates the current target epoch. \(P=2d+1\), \(Q\) and \(T_s\) denote the number of sequential epochs, number of PSG channels and number of PSG temporal samples, respectively. In this work, similar to \cite{jia2020graphsleepnet}, \(F_{de}\) is the number of the extracted differential entropy (DE) features from all PSG channels of each \(\{\bold{S}_i\}_{i=t-d}^{t+d}\). Therefore, the shape of the input data is described as \(\bold{X}=(\bold{X}_{t-d},\cdots,\bold{X}_{t},\cdots,\bold{X}_{t+d} ) \in \mathbb{R}^{P\times Q \times F_{de}}\) by concatenating the \(P\) sequential feature extracted epochs, where \(\bold{X}_i \in \mathbb{R}^{Q\times F_{de}}\) denotes the \(i\)th extracted feature epoch from the sequential epochs \(\bold{X}\). 

\textbf{Graph Signal Processing}. A graph \(\mathcal{G}_N\) can be stated as \(\{\mathcal{V},\mathcal{E},\bold{W}_N\} \), where \(\mathcal{V}\) denotes the set of graph nodes with cardinality of \(|\mathcal{V}|=N\), \(\mathcal{E}\) is the set of graph edges interpreted as node connections and \(\bold{W}_N \in \mathbb{R}^{N\times N}\) is the adjacency matrix of \(\mathcal{G}_N\) containing the edge weight \(\{\bold{W}_N (i,j)\}_{i,j=1}^N\geq0\) between the \(i\)th and \(j\)th nodes. Due to considering undirected graphs in this work, \(\bold{W}_N\) is a symmetric matrix. A signal, i.e., \(\bold{y}\in\mathbb{R}^{N\times1}\), is called a graph signal if its samples are assigned to the nodes of \(\mathcal{G}_N\). A degree matrix is a diagonal matrix \(\bold{D}_N\), which have the nodes' degrees on its diagonal. The graph Laplacian matrix \(\bold{L}_N=\bold{D}_N-\bold{W}_N\) is used to describe many useful properties of a graph; e.g., connectedness and centrality \cite{stankovic2019vertex}. The smoothness of a graph signal \(\bold{y}\) on \(\mathcal{G}_N\) is measured by the total variation (TV) of \(\bold{y}\) on \(\mathcal{G}_N\) defined as \({\textrm{TV}}_N (\bold{y})=\bold{y}^T \bold{L}_N \bold{y}= \sum_{i,j=1}^{N} {\left(\bold{y}(i)-\bold{y}(j)\right)^2 \bold{W}_N (i,j)}\). Precisely, the less the \(\textrm{TV}_N (\bold{y})\), the more similar values of \(\bold{y}\) are on connected nodes of \(\mathcal{G}_N\). The feature matrix \(\bold{Y} \in \mathbb{R}^{N \times M}\) can be considered as \(M\) graph signals \(\{\bold{y}_m \in \mathbb{R}^{N\times 1}\}_{m=1}^M\)  on \(\mathcal{G}_N\). Besides, in \(\bold{Y}\), the feature vector of the \(i\)th node can be considered as \(\Tilde{\bold{y}}_n \in \mathbb{R}^{1\times M}\), where these feature vectors can again be gathered together as \(\bold{Y}=[\Tilde{\bold{y}}_1^T|\Tilde{\bold{y}}_2^T|...|\Tilde{\bold{y}}_N^T]^T\). To investigate the overall smoothness of \(\{\bold{y}_m\}_{m=1}^M\) on \(\mathcal{G}_N\), one can measure the overall TV of these graph signals on \(\mathcal{G}_N\) as:
\begin{equation}
	\label{eq:eq1}
	\begin{split}
		\textrm{TV}_N(\bold{Y}) & =\sum_{m=1}^M{\textrm{TV}_N(\bold{y}_m)} =\sum_{i,j}^N{||\Tilde{\bold{y}}_i-\Tilde{\bold{y}}_j||_2^2\bold{W}_N(i,j)}\\
		& =2\:tr(\bold{Y}^T\bold{L}_N\bold{Y})
	\end{split}
\end{equation}

\textbf{Graph Learning}. To learn a graph from data with assumption of being smooth on the graph, Dong et al. \cite{dong2016learning} proposed to minimize Eq (\ref{eq:eq1}) w.r.t the corresponding valid graph Laplacian \(\bold{L}_N\) in a convex optimization as:

\begin{equation}
	\label{eq:eq2}
	\bold{L}_N=\underset{\bold{L}\in \mathcal{F}_N}{\arg\min}\:{tr(\bold{Y}^T\bold{L}_N\bold{Y})+\lambda||\bold{L}_N||_F^2}
\end{equation}

\noindent where \(\lambda\) balances the sides of the optimization (i.e.  the sparsest graph can be inferred with \(\lambda=0\) \cite{dong2016learning}), and  \(\mathcal{F}_N\) is the set of the valid graph Laplacians defined as:
\begin{equation}
	\label{eq:eq3}
	\begin{split}
		\mathcal{F}_N & =\{ \bold{L} \in \mathbb{R}^{N\times N} | \bold{L1}=\bold{0}, tr(\bold{L})=N, \\
		& (\forall i\neq j) \bold{L}(i,j)=\bold{L}(j,i) \leq 0 \}
	\end{split}
\end{equation}

\textbf{Product Graph Learning}. Due to the joint interaction between different domains, Kadambari et al. \cite{kadambari2020learning} showed that it is more accurate, computationally efficient and interpretable to consider the graph of interest as the Cartesian product of two significantly smaller factor graphs  \(\mathcal{G}_P\) and \(\mathcal{G}_Q\) with \(\bold{L}_P \in \mathbb{R}^{P\times P}\) and \(\bold{L}_Q \in \mathbb{R}^{Q\times Q}\) as their Laplacians, where \(PQ=N\), relevant to their corresponding domains (in this work temporal and spatial), and learn the factor graphs \(\mathcal{G}_P\) and \(\mathcal{G}_Q\) using the Kronecker sum (i.e., \(\bigoplus\)) as in \cite{hammack2011handbook}:
\begin{equation}
	\label{eq:eq4}
	\bold{L}_N=\bold{L}_P \oplus \bold{L}_Q=\bold{L}_P \otimes  \bold{I}_Q + \bold{I}_P \otimes \bold{L}_Q 
\end{equation}
\noindent where \(\bigotimes\) is the Kronecker product and \(\bold{I}_n \in \mathbb{R}^{n\times n}\) stands for the identity matrix.

Therefore, instead of having the multidomain graph signals \(\{\bold{y}_m\in \mathbb{R}^{N\times 1} \}_{m=1}^M\), the (product) graph matrices \(\{\bold{Y}_m\in \mathbb{R}^{P\times Q}\}_{m=1}^M\) represent the factor graph signals. Kadambari et al. \cite{kadambari2020learning} showed the product graph learning can be defined by \(\textrm{TV}_P (\bold{Y}_m)+\textrm{TV}_Q (\bold{Y}_m^T)\) as:

\begin{equation}
	\label{PGL_cost1}
	\begin{split}
		&\{\bold{L}_P,\bold{L}_Q\} \\
		&=\underset{\bold{L}_P\in \mathcal{F}_P, \bold{L}_Q\in \mathcal{F}_Q}{\arg\min}{\sum_{m=1}^{M}{[tr(\bold{Y}_m^T\bold{L}_P\bold{Y}_m)+tr(\bold{Y}_m\bold{L}_Q\bold{Y}_m^T)]}}\\
		&+\lambda(||\bold{L}_P||_F^2+||\bold{L}_Q||_F^2)
	\end{split}
\end{equation}

\noindent Since the goal is to learn  the sparsest graphs, therefore, using Eq. (\ref{eq:eq1}) and considering \(\lambda=0\), Eq. (\ref{PGL_cost1}) can be rewritten as:
\begin{equation}
	\label{eq:eq6}
	\{\bold{W}_P,\bold{W}_Q\}=\underset{\bold{W}_P\in \mathcal{W}_P, \bold{W}_Q\in \mathcal{W}_Q}{\arg\min}{\mathcal{L}_{\mathcal{G}_P,\mathcal{G}_Q}(\bold{W}_P,\bold{W}_Q)}
\end{equation}

\noindent where 

\begin{equation}
	\footnotesize
	\label{PGL_cost2}
	\begin{split}
		\mathcal{L}_{\mathcal{G}_P,\mathcal{G}_Q} (\bold{W}_P,\bold{W}_Q )
		= & \dfrac{1}{2}  \sum_{m=1}^{M}{[\sum_{r,s}^{P}{ ||\bold{Y}_m(r,:)-\bold{Y}_m(s,:)||_2^2 \bold{W}_P(r,s)}} \\ & + \sum_{r',s'}^{Q}{||{\bold{Y}_m(:,r')-\bold{Y}_m(:,s')||_2^2\bold{W}_Q(r',s')}}]
	\end{split}    
\end{equation}

\noindent where \(\mathcal{W}_n\) is the set of valid adjacency matrices for the undirected graphs of size \(n\) defined as:

\begin{equation}
	\label{eq:eq8}
	\mathcal{W}_n=\{\bold{W}\in \mathbb{R}^{n\times n}|\bold{W}(i,j)=\bold{W}(j,i)\geq0\}
\end{equation}

\par To jointly optimize the GL and classification tasks, the product graph learning (PGL) loss function \(\mathcal{L}_{PGL} (\bold{W}_P,\bold{W}_Q)=\mathcal{L}_{\mathcal{G}_P,\mathcal{G}_Q} (\bold{W}_P,\bold{W}_Q)\) is added to the classification loss function, i.e., Cross entropy loss.

\par \textbf{Graph Convolution}. To account the local connectivities between spatial nodes in the convolution process, the (spatial) graph signal \(\bold{s}\in \mathbb{R}^{Q\times 1}\) can be convolved with a graph Laplacian \(\bold{L}_Q \in \mathbb{R}^{Q\times Q}\) of the interested graph \(\mathcal{G}_Q\) in a more efficient manner using the Chebyshev graph convolution expansion of order \(K-1\) \cite{defferrard2016convolutional} as:

\begin{equation}
	\label{eq:eq9}
	g_{\bold{\theta}}*_\mathcal{G}\bold{s}=g_{\bold{\theta}}(\bold{L}_Q)\bold{s}=\sum_{k=0}^{K-1}{\theta_kT_k(\Tilde{\bold{L}}_Q)\bold{s}}
\end{equation}

\noindent where \(g_{\bold{\theta}}\) is the convolution kernel and \(*_\mathcal{G}\) denotes the graph convolution operator. \(\boldsymbol{\theta} \in \mathbb{R}^K\) stands for a vector containing the trainable Chebyshev coefficients. \(\Tilde{\bold{L}}_Q\) is the normalized Laplacian defined as \(\Tilde{\bold{L}}_Q=2/{\lambda_{max}}  \bold{L}_Q-\bold{I}_Q\), where \(\lambda_{max}\) is the maximum eigenvalue of \(\bold{L}_Q\). In addition, \({T}_k (x)=2x{T}_{k-1} (x)-{T}_{k-2} (x)\) denotes for the recursive Chebyshev polynomials with \(T_0 (x)=1\) and \(T_1 (x)=x\). Exploiting the approximate Chebyshev expansion, the aggregation of the information over \(K-1\) neighbor spatial nodes (i.e., \((K-1)\)-hop connections) is obtained.

\section{The Proposed ProductGraphSleepNet}
The detailed architecture if our proposed ProductGraphSleepNet is provided in Figure \ref{fig:Fig1}. The step-wise description of the different modules is detailed as:

\begin{figure*}[t!]
	\centering 
	\includegraphics[width=18cm]{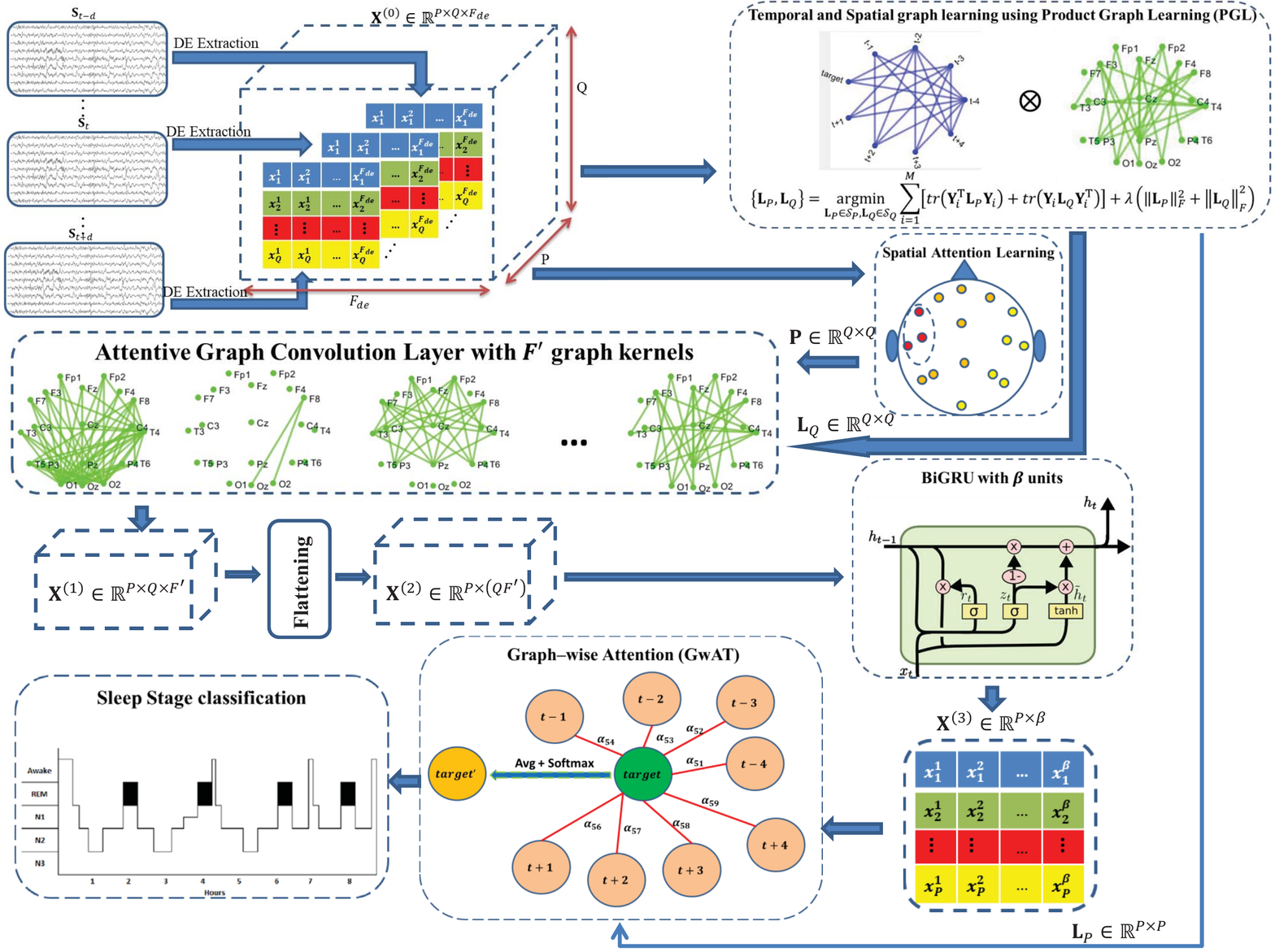} 
	\caption{The ProductGraphSleepNet. After DE feature extraction of the neighbor sleep epochs, Spatio-temporal graphs and spatial attention coefficients are learned. Then, the attentive GC layer along with BiGRU produces temporal nodes' features. Finally, using GwAT and the learned temporal graph, the final sleep staging is performed.}
	\label{fig:Fig1} 
\end{figure*}

\par \textbf{Spatial Attention Layer}. To capture the dynamic spatial attention, which adaptively changes during the transitions of the sleep stages, the spatial attention mechanism proposed by Guo et al. \cite{guo2019attention} is exploited. As shown in Figure \ref{fig:Fig1}, the input to this layer is \(\bold{X}^{(0)} \in \mathbb{R}^{P\times Q\times F_{de}}\), and the spatial attention weights, i.e., \(\bold{P}\in \mathbb{R}^{Q\times Q}\), is obtained by learning the weights \(\bold{V} \in \mathbb{R}^{Q\times Q}\), \(\bold{b}_p\in \mathbb{R}^{Q\times Q}\), \(\bold{Z}_1\in \mathbb{R}^{P\times 1}\), \(\bold{Z}_2 \in \mathbb{R}^{F_{de}\times P}\), \(\bold{Z}_3 \in \mathbb{R}^{F_{de}\times1}\) as:
\begin{equation}
	\label{eq:eq10}
	\bold{P}=\text{Softmax}(\bold{V}.\: \text{Sigmoid}(\bold{X}_{Lhs}\bold{Z}_2\bold{X}_{Rhs}^T+\bold{b}_p))
\end{equation}
\noindent where \(\bold{X}_{Lhs}=\sum_{p=1}^{P}{\bold{X}^{(0)}(p,:,:)\bold{Z}_1(p)}\in \mathbb{R}^{Q\times F_{de}}\) and \(\bold{X}_{Rhs}=\sum_{f=1}^{F_{de}}{\bold{X}^{(0)}(:,:,f)^T\bold{Z}_3(f)}\in \mathbb{R}^{Q\times P}\).

\textbf{Neural Network Modeling}. Let \(\bold{a}_{f,P}^{(i,j)}=|\bold{X}^{(0)}(i,:,f)-\bold{X}^{(0)}(j,:,f)|\in \mathbb{R}^{1\times Q}\) modeling the difference between the \(i\)th and \(j\)th sleep epochs for specified \(f\)th DE feature, and stack \(\{\bold{a}_{f,P}^{(i,j)}\}_{f=1}^{F_{de}}\) in a difference matrix \(\bold{A}_P^{(i,j)}=[{\bold{a}_{1,P}^{(i,j)}}^T |…| {\bold{a}_{F_{de},P}^{(i,j)}}^T]^T \in \mathbb{R}^{F_{de}\times Q}\). Then, the empirical difference vector \(\bold{d}_P^{(i,j)}\in \mathbb{R}^{F_{de}\times1}\) can be obtained as:

\begin{equation}
	\label{eq:eq11}
	\bold{d}_P^{(i,j)}=\dfrac{1}{Q}\sum_{q=1}^{Q}{\bold{A}_P^{(i,j)}(:,q)}
\end{equation}

\noindent  The \((i,j)\)th graph connection edge of the \(\bold{W}_P\) (i.e., \(\bold{W}_{P}(i,j)\)), which models the temporal connection between the \(i\)th and \(j\)th sleep epochs, is learned as:

\begin{equation}
	\label{eq:eq12}
	\bold{W}_{P}(i,j) = \dfrac{exp(ReLU(\bold{w}_P^T\bold{d}_P^{(i,j)}))}{\sum_{j'=1}^{P}{exp(ReLU(\bold{w}_P^T\bold{d}_P^{(i,j')}))}}
\end{equation}

\noindent with the neural network's learnable weights \(\bold{w}_P\in\mathbb{R}^{F_{de}\times1}\). Similarly, the formulation for modeling  \(\bold{W}_Q\), which contains the spatial connections between PSG electrodes and can be used for analysis of the brain connectivity, is summarized:

\begin{equation}
	\label{eq:eq16}
	\bold{a}_{f,Q}^{(i,j)}=|\bold{X}^{(0)}(:,i,f)-\bold{X}^{(0)}(:,j,f)|\in \mathbb{R}^{P\times 1}
\end{equation}

\begin{equation}
	\label{eq:eq15}
	\bold{A}_Q^{(i,j) }=[\bold{a}_{1,Q}^{(i,j)} |…| \bold{a}_{F_{de},Q}^{(i,j)}]^T \in \mathbb{R}^{F_{de}\times P}
\end{equation}

\begin{equation}
	\label{eq:eq14}
	\bold{d}_Q^{(i,j)}=\dfrac{1}{P}\sum_{p=1}^{P}{\bold{A}_Q^{(i,j)}(:,p)}
\end{equation}

\begin{equation}
	\label{eq:eq13}
	\bold{W}_{Q}(i,j) = \dfrac{exp(ReLU(\bold{w}_Q^T\bold{d}_Q^{(i,j)}))}{\sum_{j'=1}^{Q}{exp(ReLU(\bold{w}_Q^T\bold{d}_Q^{(i,j')}))}}
\end{equation}

\noindent with the learnable weights \(\bold{w}_Q\in \mathbb{R}^{F_{de}\times1}\). The output of the PGL layer are the learned spatial and temporal graph Laplacians \(\bold{L}_Q\) and \(\bold{L}_P\) based on the PGL cost function \(\mathcal{L}_{PGL}\) (\ref{PGL_cost1}) or (\ref{PGL_cost2}) mentioned in Section \ref{Prel}, as shown in Figure \ref{fig:Fig1}.

\par \textbf{Attentive Graph Convolutional Layer.} To account the spatial importance weights, local spatial connectivites in sequential epochs and as a generalization to the graph convolution of one graph signal, in this work, the \(i\)th sequential epoch \(\bold{X}^{(0)}(i,:,:)\in\mathbb{R}^{Q\times F_{de}}\), which contains \(F_{de}\) graph signals, is spatially convolved using the mentioned Chebyshev expansion and the learned spatial graph \(\bold{L}_Q\), and gets the output \(\bold{X}^{(1)}\in\mathbb{R}^{P\times Q\times F'}\), as shown in Figure \ref{fig:Fig1}, as:

\begin{equation}
	\label{eq:eq17}
	\begin{split}
		& \bold{X}^{(1)}(i,:,:) =g_{\bold{\theta}}*_{\mathcal{G}}\bold{X}^{(0)}(i,:,:)=g_{\bold{\theta}}(\bold{L}_Q)\bold{X}^{(0)}(i,:,:) \\
		& =\sum_{k=0}^{K}{{\left[T_k(\Tilde{\bold{L}}_Q)\odot \bold{P}\right]}\bold{X}^{(0)}(i,:,:)\bold{\Theta}^{(k)}};  \:\:i=1,2,..,P     
	\end{split}
\end{equation}

\noindent where \(\bold{\Theta}^{(k)}\in\mathbb{R}^{F_{de}\times F'}\) is the \(k\)th trainable Chebyshev coefficients matrix, \(\odot\) denotes the Hadamard (element-wise) product, and \(\bold{P}\) is learned from the spatial attention layer. 

\textbf{BiGRU.} The output \(\bold{X}^{(1)}\in\mathbb{R}^{P\times Q\times F'}\) is then flattened to obtain feature vectors \(\bold{X}^{(2)}\in\mathbb{R}^{P\times (QF')}\) in Figure \ref{fig:Fig1}, and fed to a BiGRU module with \(\beta\) units to learn the sleep stage transition rules between these \(P\) sequential epochs. Therefore, the BiGRU module gets output \(\bold{X}^{(3)}\in\mathbb{R}^{P\times\beta}\). This obtained \(\bold{X}^{(3)}\) in Figure \ref{fig:Fig1} is considered as the feature matrix of the nodes of the learned temporal graph \(\bold{L}_P\) and is fed to the GwAT module. 

\par \textbf{Graph-wise Attention Network (GwAT).} To account the importance weights of the different neighbor sleep epochs and also learn attentive temporal feature vectors, we modify the celebrated Graph Attention Network (GAT) so it can be used in graph classification scenarios. Considering \(\bold{X}^{(3)}\in\mathbb{R}^{P\times\beta}\) and \(\bold{W}_P\) be the feature matrix and adjacency matrix of the learned temporal graph \(\mathcal{G}_P\) to our network, the goal here is learning attention coefficients \(\pmb{\alpha} \in\mathbb{R}^{P\times P}\), and obtaining new attentive feature vectors \(\bold{X}^{(4) }\in\mathbb{R}^{P\times F_{\textit{GwAT}}}\) as:

\begin{equation}
	\label{eq:eq18}
	\pmb{\alpha}(r,s)=\dfrac{\hat{\pmb{\alpha}}(r,s)}{\sum_{s'=1}^{P}{\hat{\pmb{\alpha}}(r,s')}}
\end{equation}


\begin{equation}
	\label{eq:eq19}
	\begin{split}
		& \hat{\pmb{\alpha}}(r,s)= \bold{W}_P(r,s)\times\\
		& \exp\left(\text{LeakyReLU}\left(\pmb{\gamma}^T\left[\bold{X}^{(3)}(r,:)\bold{W}||\bold{X}^{(3)}(s,:)\bold{W}\right]\right)\right)
	\end{split}
\end{equation}

\noindent where \(\boldsymbol{\gamma}\in\mathbb{R}^{2F_{\textit{GwAT}}\times1}\) and \(\bold{W}\in\mathbb{R}^{\beta\times F_{\textit{GwAT}}}\) are the learnable weights of the GwAT module, and \(||\) stands for the concatenation operator. Moreover, the new attentive feature vectors of the temporal graph nodes can be obtained as:

\begin{equation}
	\label{eq:eq20}
	\bold{X}^{(4)}=\pmb{\alpha}\bold{X}^{(3)}\bold{W}
\end{equation}

To increase the learning capacity, one can have \(K_{\textit{GwAT}}\) attention heads in which the mentioned equations can be briefly indexed w.r.t the \(k\)th (\(k=1,...,K_{\textit{GwAT}}\)) attention head:

\begin{equation}
	\label{eq:eq18}
	\pmb{\alpha}^{(k)}(r,s)=\dfrac{\hat{\pmb{\alpha}}^{(k)}(r,s)}{\sum_{s'=1}^{P}{\hat{\pmb{\alpha}}^{(k)}(r,s')}}
\end{equation}

\begin{equation}
	\label{eq:eq22}
	\begin{split}
		& \hat{\pmb{\alpha}}^{(k)}(r,s)= \bold{W}_P(r,s)\times\\
		& \exp\left(\text{LeakyReLU}\left(\pmb{\gamma}^T\left[\bold{X}^{(3)}(r,:)\bold{W}^{(k)}||\bold{X}^{(3)}(s,:)\bold{W}^{(k)}\right]\right)\right)
	\end{split}
\end{equation}

\begin{equation}
	\label{eq:eq23}
	{\bold{X}^{(4)}}^{(k)}=\pmb{\alpha}^{(k)}\bold{X}^{(3)}\bold{W}^{(k)}
\end{equation}

\noindent Finally, the GwAT's classification output \(\bold{X}_o\) is defined as:

\begin{equation}
	\label{eq:eq25}
	\bold{X}_o=\sigma(\dfrac{1}{K_{\textit{GwAT}}}\sum_{k=1}^{K_{\textit{GwAT}}}{(\dfrac{1}{P}\sum_{p=1}^{P}{{\bold{X}^{(4)}}^{(k)}(p,:)})}) \in \mathbb{R}^{1\times {F_{\text{\textit{GwAT}}}}}
\end{equation}

\noindent where \(\sigma\) is the \textit{softmax} function. Note that in this work, \(F_{\textit{GwAT}}=5\) is equal to the number of sleep stages.

\section{Experimental Setup}

\noindent This work focuses on developing an automated interpretable sleep staging classifier. To evaluate the proposed ProductGraphSleepNet, two well-known public datasets are used. 

\textbf{Dataset}. 1) The MASS-SS3 database \cite{o2014montreal} contains PSG signals (with sampling frequency 256 Hz) of 62 healthy subjects recorded on 20 EEG, 3 electromyograms (EMG), 2 electrooculograms (EOG), and 1 electrocardiogram (ECG) electrodes. The sleep epochs were staged based on five sleep stages, i.e., W, REM, N1-3, by sleep staging specialists following the AASM standard \cite{iber2007aasm}. 2) The SleepEDF dataset \cite{goldberger2000physiobank,kemp2000analysis} includes data from 20 healthy subjects, 2 EEG, 1 EOG, 1 EMG, and 1 oro-nasal respiration channels with the sampling frequency of 100 Hz, where are scored by R\&K standard. To have similar labeling space (five-stage), the sleep stages of SleepEDF are converted to the AASM standard. For SleepEDF dataset, only the available EEG and EOG channels are used.

\textbf{Preprocessing}. Firstly, each of the PSG channels was decomposed into sub-frequency bands 0.5-4 Hz, 2-6 Hz, 4-8 Hz, 6-11 Hz, 8-14 Hz, 11-22 Hz, 14-31 Hz, 22-40 Hz, and 31-50 Hz. Then, without any further pre-processing or denoising procedures, DE features were extracted from them. Thus, in this work, \(Q=26\) and \(Q=3\) for the MASS-SS3 and SleepEDF datasets, respectively, and \(F_{de}=9\). Besides, due to taking into account four epochs before and after the target sleep epoch, we have \(d=4\) and \(P=2d+1=9\).

\textbf{Training Setup.} We conduct our experiment on the MASS-SS3 dataset using a 16-fold cross-validation scheme, in which each of the first 15 folds contains four subjects' data, while the last fold consists data from two subjects. As validation data for each test fold, one fold is randomly selected, and the best-trained model on this validation data is used to apply on the unseen test fold. For SleepEDF, we use the Leave One Subject Out (LOSO) cross validation (with the randomly selected one subject as validation data for each test subject). The values of the hyperparameters in the model are as follow: $F'=10$, $\beta=256$, $K_{\text{GwAT}}=20$, $\text{Number of training epochs} = 100$, $\text{Chebyshev Order K} = 3$, $\text{Dropout probability}= 0.6$, $\text{Batchsize}=1024 $, $\text{Learning rate} = 0.001$, $\text{Optimizer=Adam}$ \cite{kingma2014adam}. All the training steps are implemented using Tensorflow \cite{abadi2016tensorflow}.

\section{Experimental Results and Discussion}
To evaluate the effectiveness of ProductGraphSleepNet, some ablation baselines and also some recent relevant studies that reported their performance metrics on the MASS-SS3 and SleepEDF datasets are considered.

\textbf{Ablation study.} To investigate the effectiveness of the proposed modules in our ProductGraphSleepNet, we implemented two major baselines on the MASS-SS3 dataset, named Baseline 1 and 2 in Table \ref{MASS-SS3} with the details as follows: Baseline 1: Jia et al. \cite{jia2020graphsleepnet} proposed a graph neural network for performing sleep staging tasks. In the Baseline 1, the temporal information are simply concatenated, and the BiGRU is not exploited. Baseline 2: To investigate the impact of the GwAT in the proposed method, the GwAT is replaced with the simple concatenation of the outputs of the BiGRU module.

\begin{figure}[t!]
	\centering
	\subfloat[]{{\includegraphics[width=.43\textwidth, trim=4.5cm 1cm 5.5cm 2.5cm]{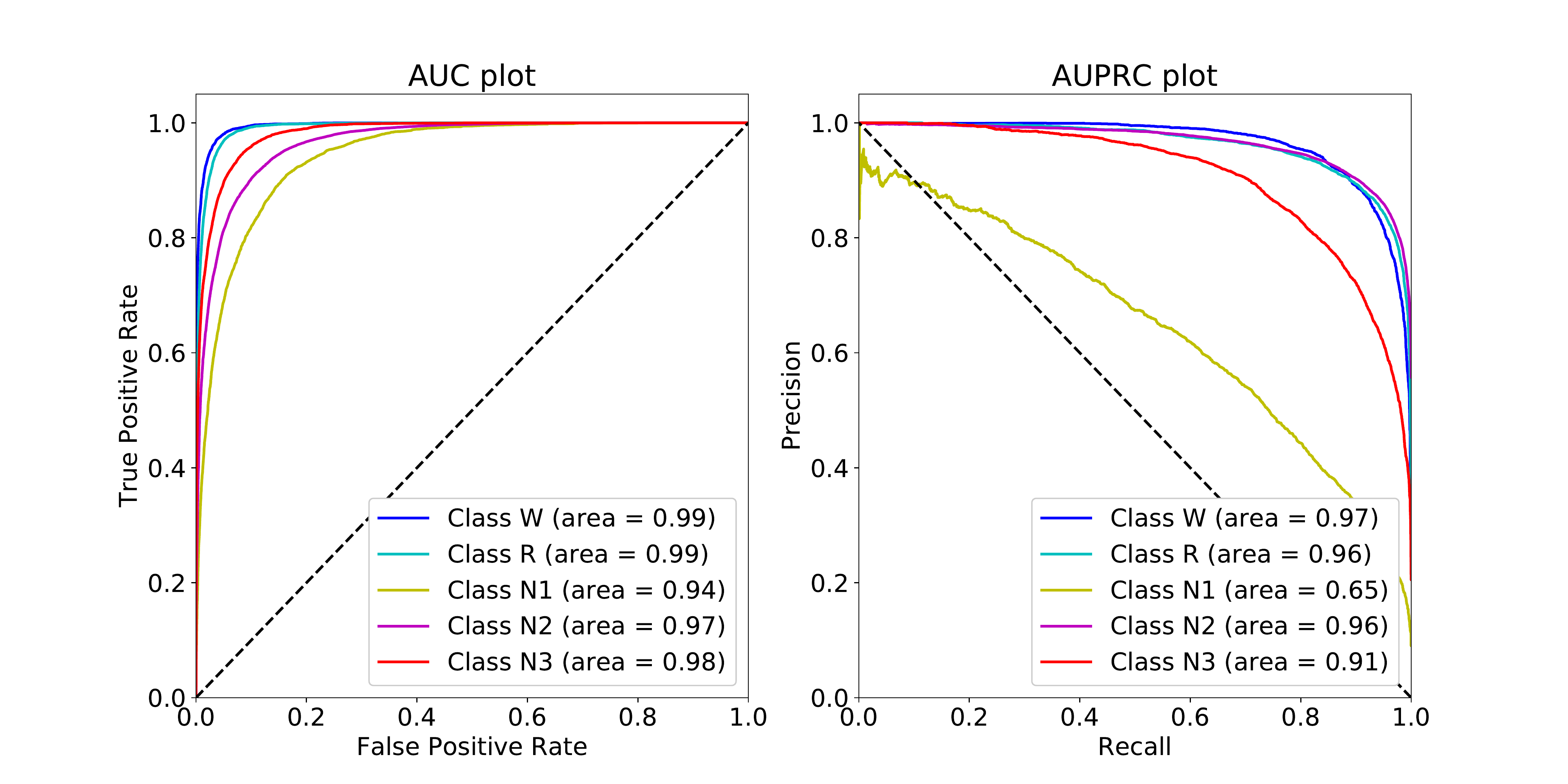} }\label{fig:mycaption-a}}
	
	\subfloat[]{{\includegraphics[width=.43\textwidth, trim=4.5cm 1cm 5.5cm 2cm]{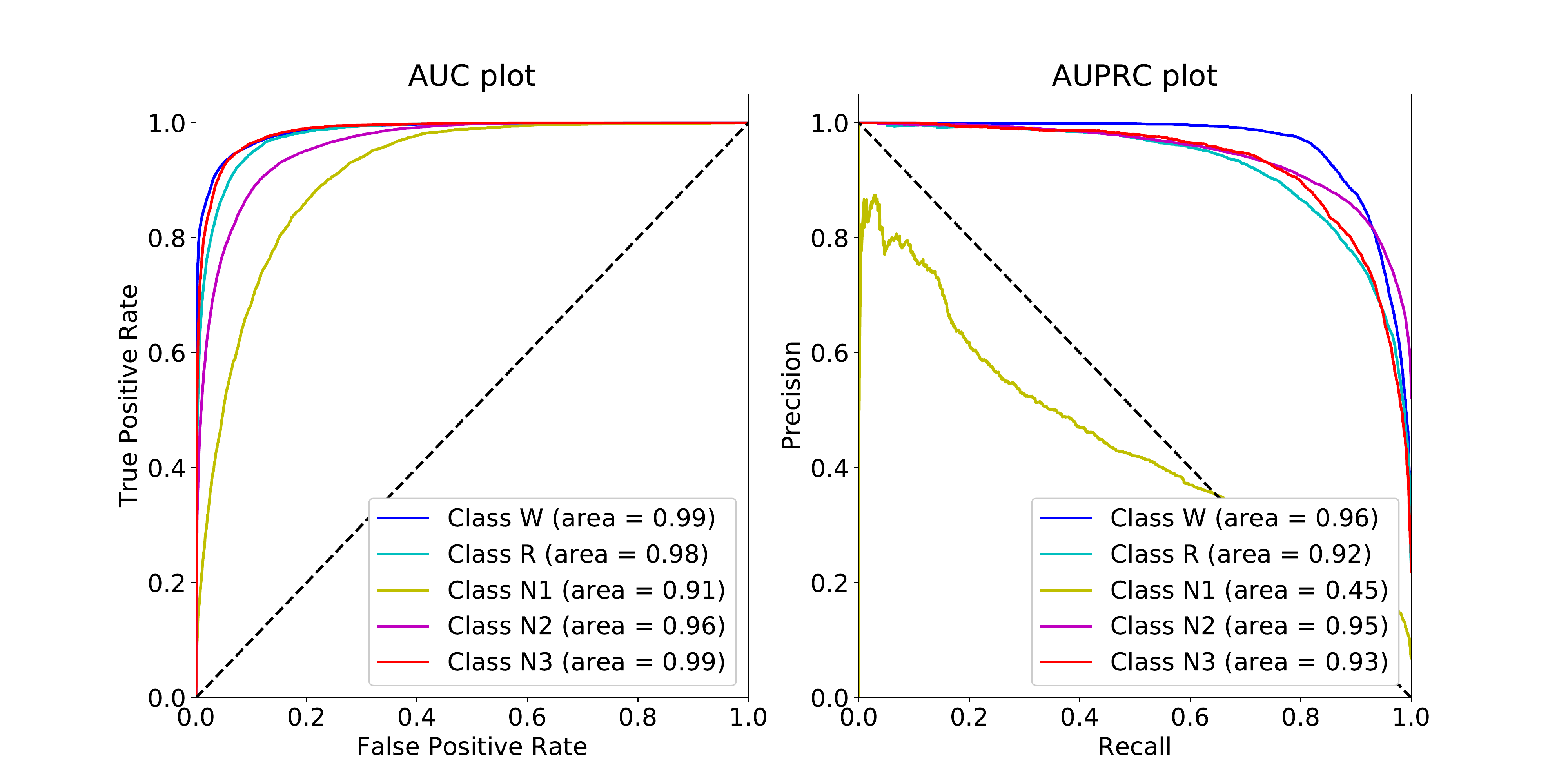} }\label{fig:mycaption-b}}
	
	\caption{AUC/AUPRC of the sleep staging task obtained using the proposed method: a) MASS-SS3 and b) SleepEDF.}
	\label{AUC_AUPRC_Confusion}
\end{figure}

\begin{figure}[t!]
	\centering
	\subfloat[]{{\includegraphics[width=6.5cm, trim=0cm 0cm 0cm 0cm]{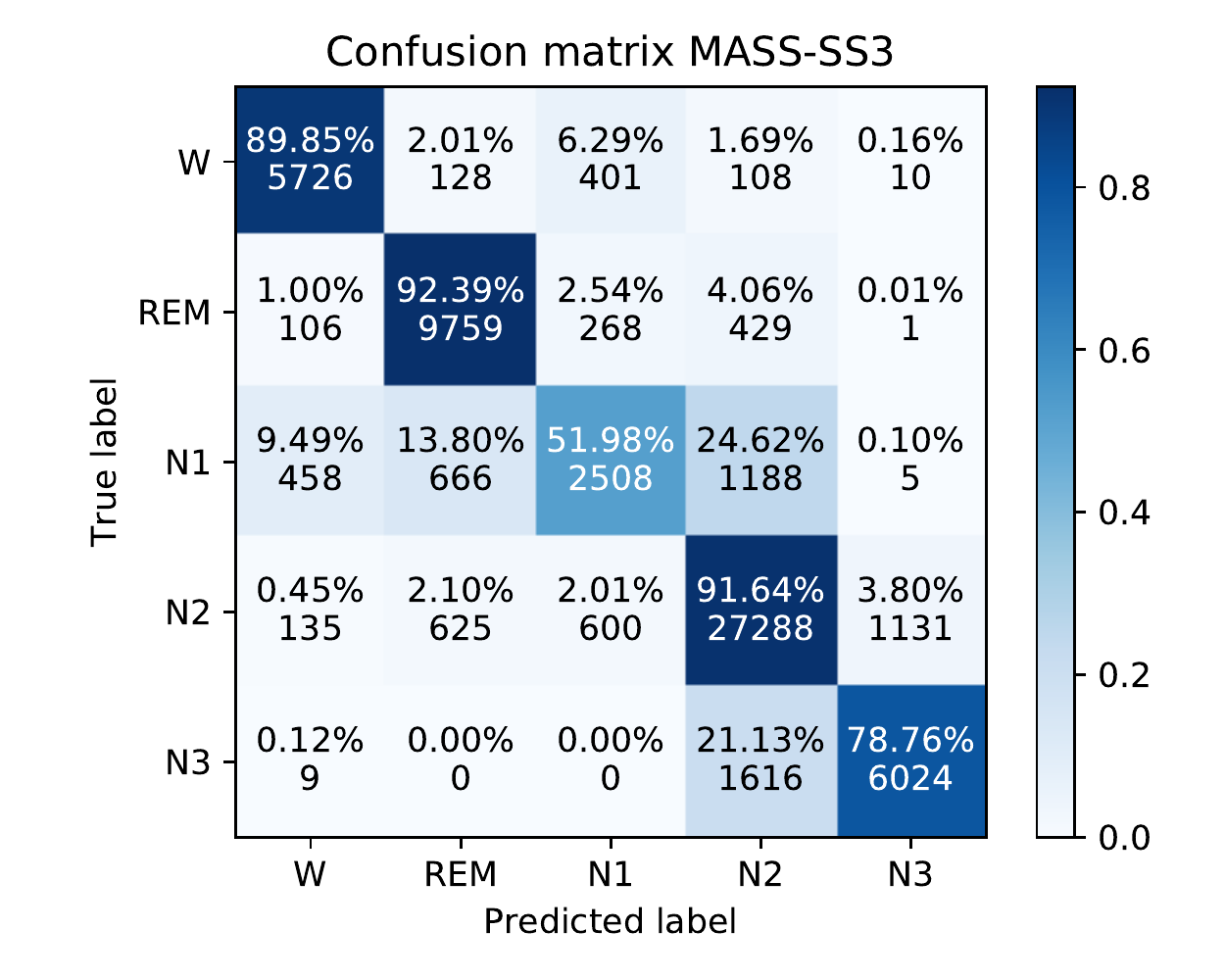} }}%
	\qquad
	\subfloat[]{{\includegraphics[width=6.5cm, trim=0cm 0cm 0cm 0cm]{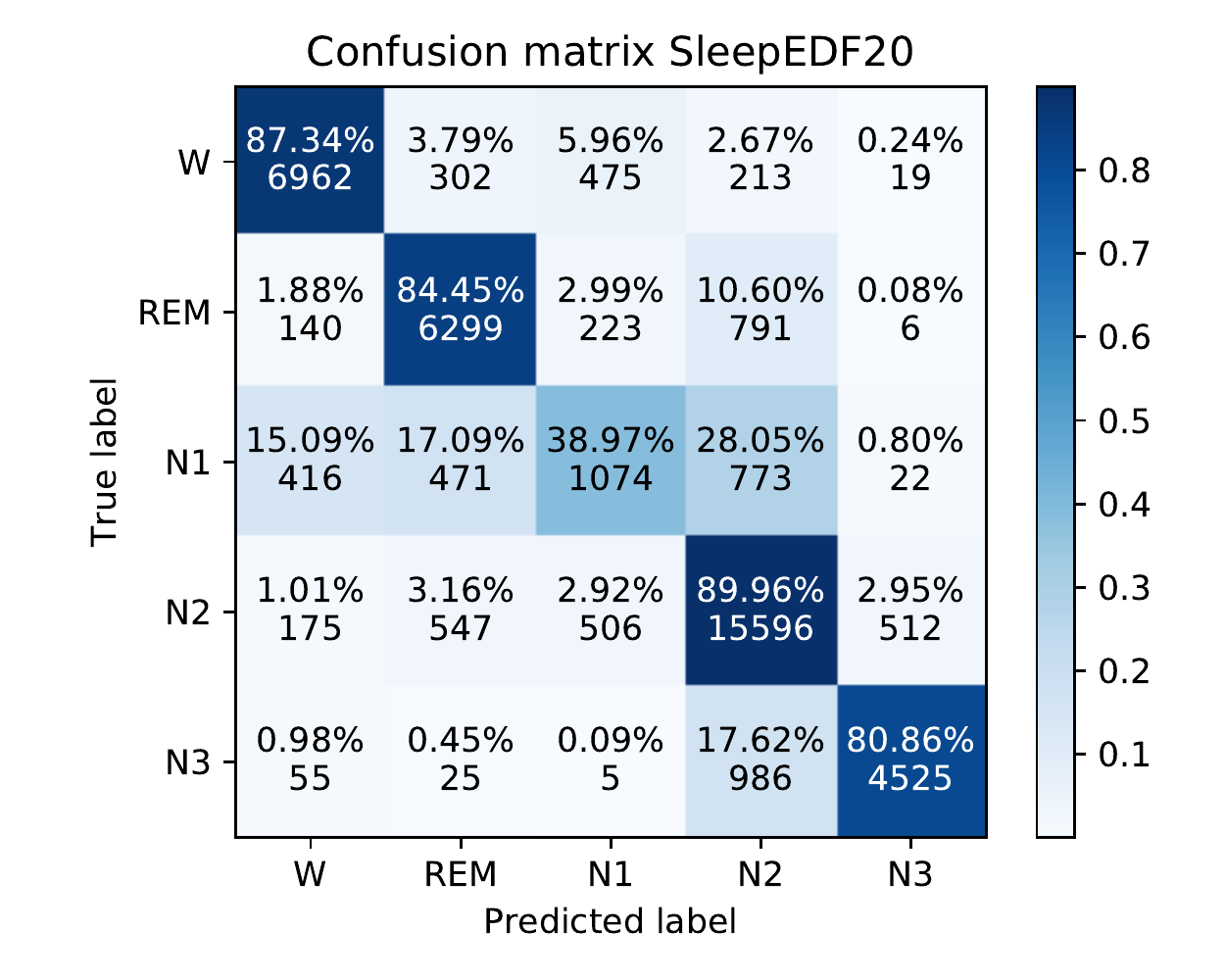}}}%
	\caption{Confusion matrix for the sleep staging task using the proposed method: a) MASS-SS3 and b) SleepEDF.}
	\label{ConfusionMatrix}%
\end{figure}

\textbf{Comparison with State-of-the-Art Methods.} The considered studies for MASS-SS3 are: 1) Dong et al. \cite{dong2017mixed} provides a comprehensive study in which the performance of the traditional Random Forest (RF) and Support Vector Machine (SVM) classifiers were compared with a mixed neural network, which is a hybrid of Multi-Layer Perceptron (MLP) and Long Short Term Memory (LSTM) modules; 2) Supratak et al. \cite{supratak2017deepsleepnet} considered a hybrid model of the CNN and BiLSTM modules, which take into account the neighbor sleep epoch data to learn the transition rules between the sleep stages; 3) Chambon et al. \cite{chambon2018deep} performed a study that handles the multimodality of the PSG signals in a temporal sleep staging scheme;  4) Jiang et al. \cite{jiang2019robust} proposed a Hidden Markov Model (HMM)-based robust classifier that handles multimodal PSG learning; 5) Sun et al. \cite{sun2019hierarchical}: Learning comprehensive and sequence feature learning using a hierarchical neural network based on a hybrid combination of the CNN and BiLSTM networks; 6) Phan et al. \cite{phan2019seqsleepnet} proposed an attentive hierarchical recurrent neural network to process the sequences of the sleep epochs. Similarly, for SleepEDF evaluation, the considered studies (as well as the described \cite{supratak2017deepsleepnet} study), are: 1) Tsinalis et al. \cite{tsinalis2016automatic} investigated the impact of the different time-frequency features for using as input data to a DL network. 2) Vilamala et al. \cite{vilamala2017deep} analyzed the combination of the Multitaper spectral with a CNN. 3) Seo et al. \cite{seo2020intra} exploited BiLSTM network for learning Intra/Inter-epoch temporal dependencies. Table \ref{MASS-SS3} and Table \ref{SleepEDF} provide the comparisons among these techniques on the MASS-SS3 and SleepEDF datasets, respectively, where bold and underlined metrics are corresponds to the best and second best performances, respectively. Based on these results, one can conclude that our proposed network has competitive performance compared to the state-of-the-art methods with not significant performance difference rather than the best performances (even with the small number of channels in the SleepEDF dataset), and more importantly our proposed method provides a medically interpretable automated sleep staging algorithm. Moreover, results of the baseline 1 show that capturing the dynamics of the sleep stage transition using a temporal module (e.g., BiGRU, PGL, and GwAT) plays a critical role in the sleep staging task. The results obtained using the baseline 2 show that GwAT increases the discriminability and transferability of the model. Area Under Curve (AUC), Area Under Precision-Recall Curve (AUPRC) plots, and the confusion matrix using ProductGraphSleepNet on the MASS-SS3 and SleepEDF datasets, admitting the great discriminability of the proposed method, are shown in Figures \ref{AUC_AUPRC_Confusion} and \ref{ConfusionMatrix}, respectively. 

\begin{table*}[hbt!]
	\centering
	\caption{Performance (i.e., Overall Accuracy, F1-score and Kappa, as well as per class F1-score) comparison between the ProductGraphSleepNet and state-of-the-art on MASS-SS3, where bold and underlined metrics are corresponds to the best and second best performances, respectively.}
	\begin{tabular}{||c |c| c| c| c| c| c| c| c|c||} 
		\hline
		& Method & Accuracy & F1-score & Kappa & Wake & REM & N1 & N2 & N3  \\ [0.5ex] 
		\hline\hline
		Baseline 1 & Modified GraphSleepNet & 0.845 & 0.792 & 0.773 & 0.878 & 0.874 & 0.527 & 0.889 & 0.788  \\ \hline
		Baseline 2 &  Simple concatenation & 0.853 & 0.801 & 0.782 & 0.889 & 0.887 & 0.555 & 0.894 & 0.782 \\ \hline
		Dong et al. \cite{dong2017mixed} & SVM & 0.797 & 0.750 & - & 0.786 & 0.792 & 0.487 & 0.861 & \textbf{0.825} \\ \hline
		Dong et al. \cite{dong2017mixed} & RF & 0.817 & 0.724 & - & 0.782 & 0.794 & 0.351 & 0.880 & \underline{0.815} \\ \hline
		Dong et al. \cite{dong2017mixed} & MLP+LSTM & 0.859 & 0.805 & - & 0.846 & 0.861 & 0.563 & \textbf{0.907} & 0.848 \\ \hline
		Supratak et al. \cite{supratak2017deepsleepnet} & DeepSleepNet & 0.862 & 0.817 & 0.800 & 0.873 & 0.893 & \textbf{0.598} & 0.903 & \underline{0.815}\\ \hline
		Chambon et al. \cite{chambon2018deep} & CNN & 0.739 & 0.673 & 0.640 & 0.730 & 0.764 & 0.294 & 0.812 & 0.765 \\ \hline
		Jiang et al. \cite{jiang2019robust} & RF+HMM & 0.808 & 0.793 & 0.710 & - & - & - & - & - \\ \hline
		Perslev et al. \cite{perslev2021u} & U-Sleep & - & 0.800 & - & \textbf{0.930} & \textbf{0.910} & 0.540 & 0.870 & 0.750 \\ \hline
		Phan et al. \cite{phan2019seqsleepnet} & SeqSleepNet & \textbf{0.871} & \textbf{0.833} & \textbf{0.815} & - & - & - & - & - \\ \hline
		Our Method & ProductGraphSleepNet & \underline{0.867} & \underline{0.818} & \underline{0.802} & \underline{0.894} & \underline{0.898} & \underline{0.583} & \underline{0.904} & 0.813 \\ \hline
		\hline
	\end{tabular}
	\label{MASS-SS3}
\end{table*}

\begin{table*}[hbt!]
	\centering
	\caption{Performance (i.e., Overall Accuracy, F1-score and Kappa, as well as per class F1-score) comparison between the ProductGraphSleepNet and state-of-the-art on SleepEDF, where bold and underlined metrics are corresponds to the best and second best performances, respectively.} 
	\begin{tabular}{||c |c| c| c| c| c| c| c| c|c||} 
		\hline
		& Method & Accuracy & F1-score & Kappa & Wake & REM & N1 & N2 & N3  \\ [0.5ex] 
		\hline\hline
		Tsinalis et al. \cite{tsinalis2016automatic} & Time-frequency features & 0.748 & 0.698 & 0.65 & 0.437 & 0.654 & 0.806 & 0.849 & 0.745\\ \hline
		Supratak et al. \cite{supratak2017deepsleepnet} & DeepSleepNet & 0.820 & 0.769 & 0.760 & 0.847 & 0.824 & \underline{0.466} & \textbf{0.898} & 0.848 \\ \hline
		Vilamala et al. \cite{vilamala2017deep} & Multitaper spectrals + CNN & 0.813 & 0.765 & 0.740 & 0.809 & 0.819 & \textbf{0.474} & 0.862 & \underline{0.862} \\ \hline
		Seo et al. \cite{seo2020intra} & Intra-/inter-epoch BiLSTM & \textbf{0.839} & \textbf{0.776} & \textbf{0.78} & \underline{0.877} & \underline{0.825} & 0.434 & \underline{0.877} & \textbf{0.867}\\ \hline
		Our Method & ProductGraphSleepNet & \underline{0.838} & \underline{0.774} & \underline{0.775} & \textbf{0.886} & \textbf{0.834} & 0.426 & 0.874 & 0.847 \\ \hline
		\hline
	\end{tabular}
	\label{SleepEDF}
\end{table*}

\begin{figure*}[t!]
	\centering 
	\includegraphics[width=18cm]{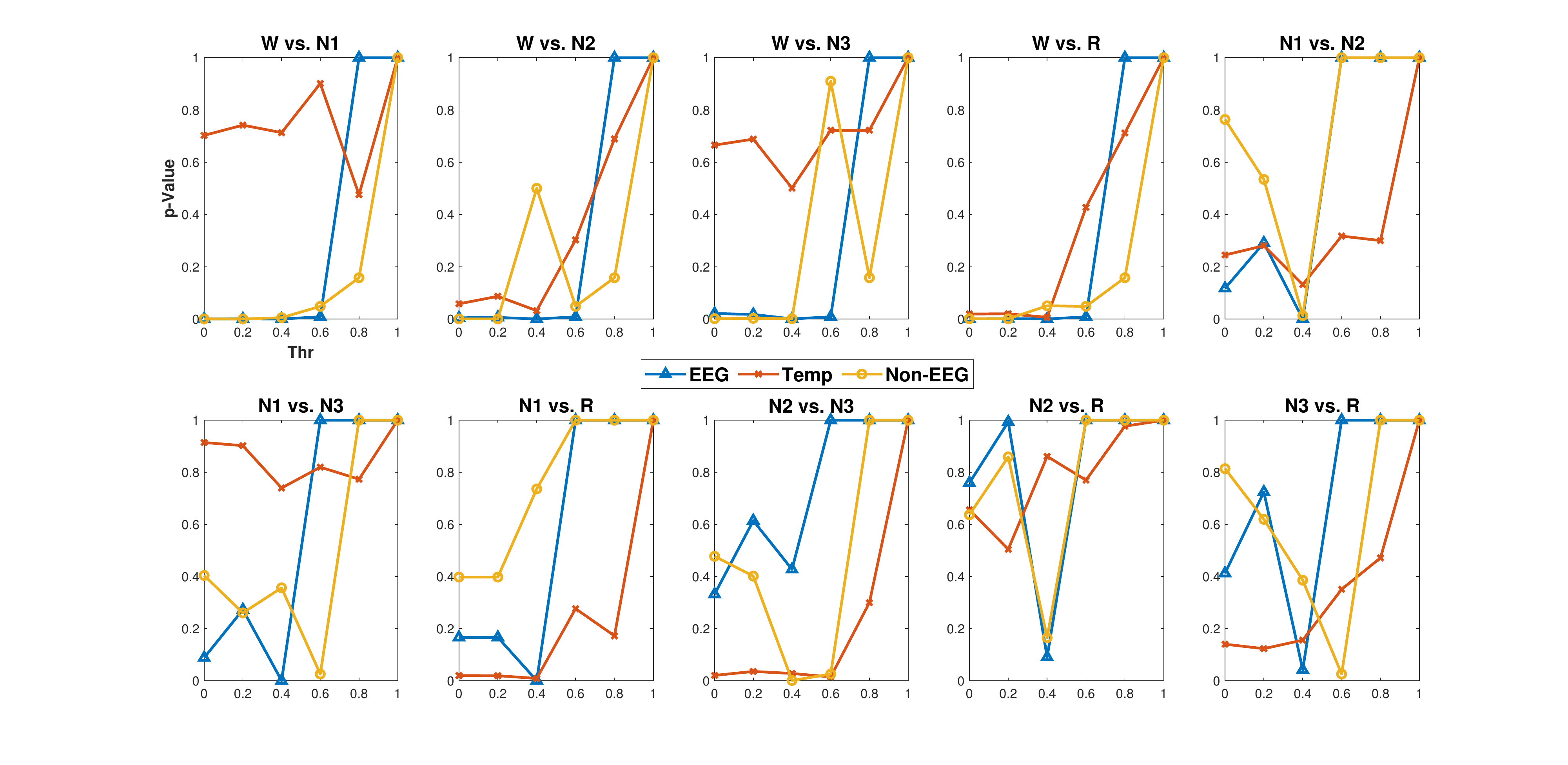} 
	\caption{The calculated \(p\)-values from \(t\)-test statistical approach for EEG vs. EEG (EEG in this figure), EEG vs. Non-EEG (Non-EEG in this figure), and also temporal sleep epochs graph connectivities (Temp in this figure), across the values of \(\{0, 0.2, 0.4, 0.6, 0.8, 1\}\) for each pair of sleep stages.}
	\label{PairwisePValues} 
\end{figure*}

\textbf{Analysis of the Learned Spatial/Temporal Graphs on
	the MASS-SS3 dataset.} Analysis of the functional brain connectivity through obtained spatial brain graphs, e.g., variation of the functional connectivities within sleep stage transitions \cite{nguyen2018exploring}, has been addressed in pioneer neuroscientific literature as an active research area \cite{alper2013weighted, huang2018graph}. In the following, the learned temporal and spatial graphs from the proposed ProductGraphSleepNet corresponding to EEG vs. EEG and EEG vs. Non-EEG electrodes and averaged over class samples are analyzed to gain insights about the brain pattern transitions and illustrate the interpretability aspect of the proposed network. Firstly, the learned graphs need to be binarized according to a well-specified threshold, which we devise and adopt a statistical approach to obtain a specific threshold discriminating the obtained graphs the most. In this way, we considered each pair of graphs corresponding to the sleep stages and vary the edge values of these graphs across the span of  \(\{0, 0.2, 0.4, 0.6, 0.8, 1\}\). Then, the \(p\)-values obtained from the \(t\)-test approach are calculated as depicted in Figure \ref{PairwisePValues}, and the lowest (non-zero) \(p\)-value is selected to illustrate the most discriminating patterns, i.e., the most statistical significance, in the obtained graphs. Here, \(Thr=0.4\) is selected and used for the binarization of the obtained binary graphs.    

\begin{figure}[t!]
	\centering 
	\includegraphics[width=9cm]{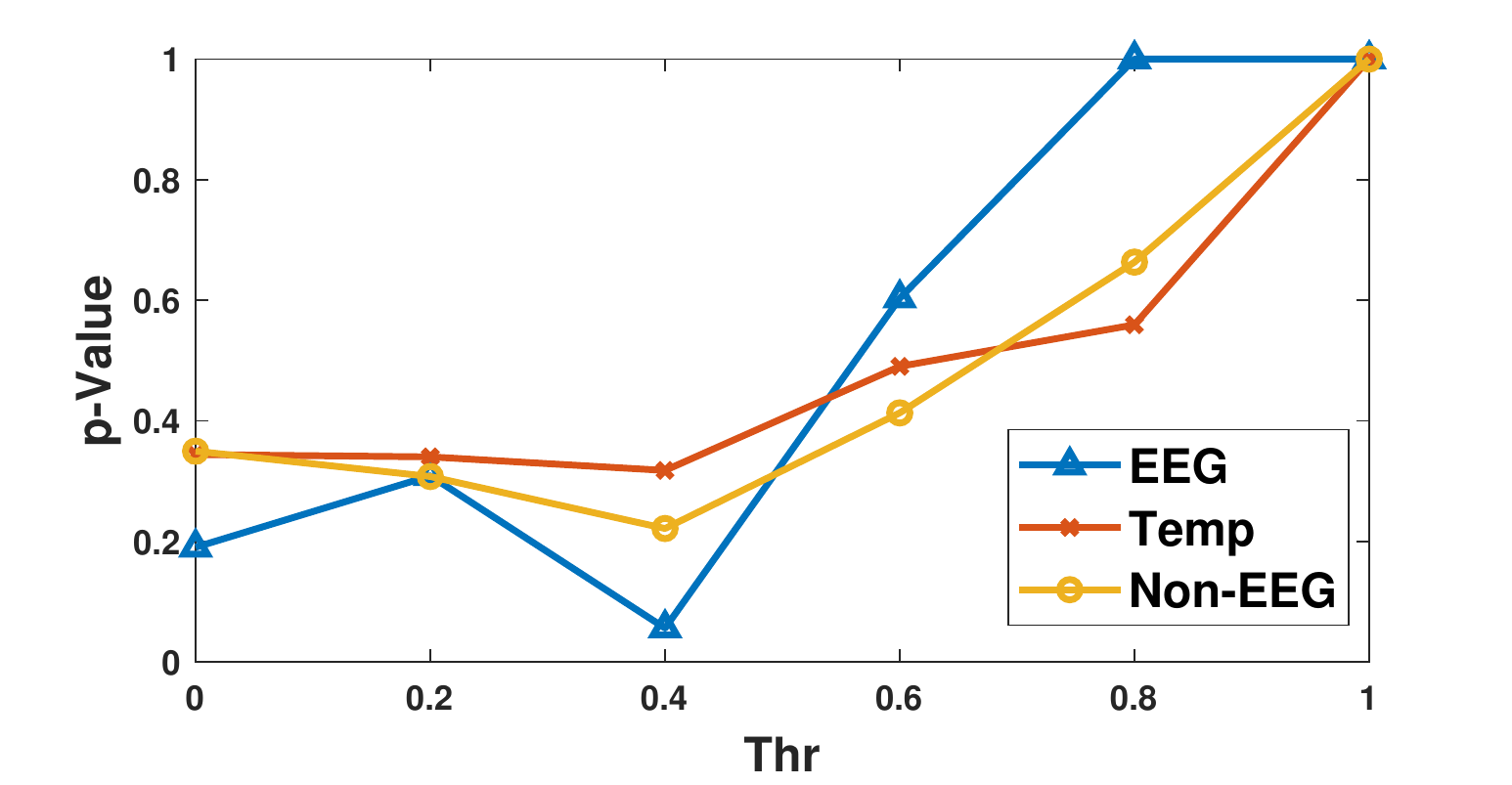} 
	\caption{The \(p\)-values obtained from \(t\)-test and averaged over the pairwise sleep stage plots for EEG vs. EEG (EEG in this figure), EEG vs. Non-EEG (Non-EEG in this figure), and also temporal sleep epochs graph connectivities (Temp in this figure), in the span of \(\{0, 0.2, 0.4, 0.6, 0.8, 1\}\).}
	\label{AvgPValues} 
\end{figure}

\begin{figure*}[hbt!]
	\centering 
	\includegraphics[width=14cm, trim=7cm 1cm 5cm 0.9cm]{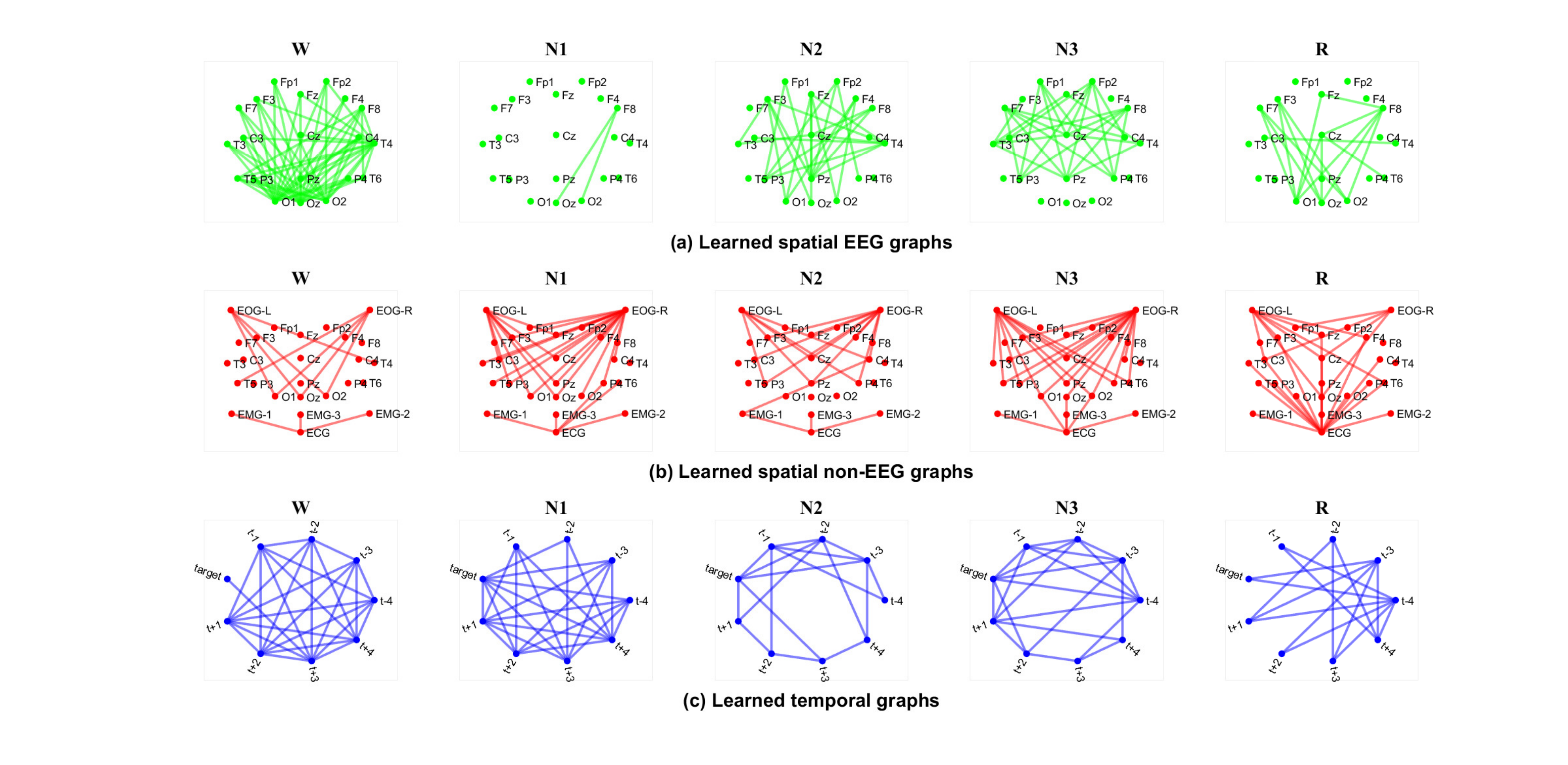} 
	\caption{The binarized learned spatio-temporal graphs corresponding to different sleep stages from PSG recordings of MASS-SS3 dataset and averaged over the class samples. a) The significant conncetivity reduction of the non-Wake stages vs. Wake is shown in this figure. b) The notable increased connectivity between ECG and EEG channels in REM stage compared to the others is illustrated.}
	\label{Fig3} 
\end{figure*}

The binarized learned spatial graphs implying the connections across the EEG electrodes have been depicted in Figure \ref{Fig3} (a). It can be seen that the overall functional connectivities increase in the Wake state compared to other Non-REM ones. This observation is compatible with mentioned neuroscientific findings implying the reduction in hypothalamic functional connectivity of the Non-REM stages for (probably) stabilizing sleep \cite{kaufmann2006brain, tarun2021nrem}. Besides, it can be observed that the occipital region is considerably more active during the REM stage compared to the other Non-REM ones, which is supported by the mentioned and observed higher occipital metabolism in REM compared to Non-REM ones leading to more connectivity \cite{kjaer2002regional, kaufmann2006brain}.

Figure \ref{Fig3} (b) shows the binarized averaged learned connections between non-EEG channels (i.e., EOG Left, EOG Right, EMG chin1-3, and ECG) and EEG channels. For clarity, the connections between the EEG channels are ignored in this figure. Based on these learned connections, one can conclude that: 1) There are significantly more connections between ECG and EEG channels during REM compared to the other stages, especially NREM. This is consistent with \cite{varoneckas1986components, penzel2003dynamics}, where they showed that the increased activity of the heart rate also its connections stem from the increased influence of the brain on the autonomous nervous system. 2) The connection between EOG and EEG channels has a discriminative pattern between the sleep stages of NREM, REM, and wake, which aligns with the results of \cite{virkkala2007automatic}, where they detected discriminative synchronized EEG activity in the Wake and REM vs. the other stages by measuring the peak-to-peak amplitude and cross-correlation in the 0.5-6 Hz band between two EOG channels.

The binarized learned temporal graphs averaged over the class samples and corresponding to sleep stages are depicted in Figure \ref{Fig3} (c). As can be seen in this figure, these graphs have not necessarily had a tree-like topology, which means it is not efficient to assume that only the one-hop neighbor temporal epochs are connected. In fact, considerable variant types of connections are observed in this figure, such as the connection between the first (i.e., \(t-4\)) and the fifth (i.e., \(t+1\)) epoch corresponding to the REM stage, which is well-discriminative across other stages helping to probably more efficient classification. Note that these kinds of connections have been ignored in related studies \cite{jia2020graphsleepnet} by simply concatenation of the obtained feature vectors from the neighbor epochs.

\section{Conclusion}
This paper proposed an adaptive and automatic sleep staging network, namely ProductGraphSleepNet, which exploits Product Graph Learning (PGL) along with a Graph Convolutional Neural Network (GCN) to learn Spatio-temporal graphs obtained from the neighbor (sequential) sleep epochs. Due to the non-Euclidean nature of brain signals, using a graph-based structure preserves the geometry information. In particular, the proposed method learns the sequential connections between neighbor sleep epochs, which capture the dynamics of sleep stage transitions. The experimental results on the two public datasets, Montreal Archive of Sleep Studies (MASS) SS3 and  SleepEDF datasets, show that our method is competitive with state-of-the-art methods, as well as being medically interpretable in, contrast to the current automatic sleep staging methods. 

\bibliographystyle{unsrt}
\bibliography{References}

\end{document}